\documentstyle[preprint,prc,aps]{revtex}
\begin{document}
\draft
\title
{\bf \LARGE Caloric curve in Au + Au collisions}
\narrowtext
\author{C. B. Das and L. Satpathy}
\address{Institute of Physics, Bhubaneswar - 751 005, India}
\maketitle
\begin{abstract}
Realistic caloric curves are obtained for $^{197}Au+^{197}Au$ reaction 
with incident energy ranging from 35 to 130 MeV/nucleon in the dynamic
statistical multifragmentation model. It is shown that
for excitation energy 3 to 8 MeV/nucleon, the temperature remains
constant in the range 5 to 6 MeV, which is close to experiment.
The mechanism of energy deposition through the tripartition of the
colliding system envisaged in this model together with 
inter-fragment nuclear interaction are found to play important role.
A possible signature of liquid-gas phase transition is seen in the
specific heat distribution calculated from these caloric curves, and
the critical temperature is found to be $\sim$ 6 to 6.5 MeV.
\end{abstract}
\narrowtext
\newpage

It has been speculated more than ten years ago that the nuclear system
will show liquid-gas phase transition. This is based on two well known
facts, namely, (i) the similarity of the nucleon-nucleon interaction 
with its general feature of repulsion followed by attraction with the 
Van der Waal force, and (ii) the overwhelming success of the liquid drop 
model. The earliest search in this regard has been in the high energy
proton induced $Xe$ and $Kr$ reaction using the prescription of Fisher
droplet model\cite{Fi67}. The successful description of the mass yield 
characterized by power law distribution, through such a
model was considered indicative of the signature of liquid-gas phase 
transition. However, many other models without having explicitly the 
mechanism of liquid-gas phase transition in them [2-5], could also 
explain the data. This has dampened the interest of the community 
about this interesting possible phenomenon. However,
in the last couple of years there has been renewed interest due to more 
extensive experimental investigation\cite{Cc83,Jf86}
 to find critical exponents in the
multifragmentation of $Au$ nuclei. A desirable feature of any experimental 
detection of a phenomenon should be the measurement of such observables
whose interpretation would require minimal amount of theory or model.
This would lead to ``theory independence'' of the conclusions.
In the present context, a more appropriate attempt would be to measure
the excitation energy and temperature of hot nuclear systems.
The caloric curve thus obtained, should show the well known feature
of liquid-gas phase transition in a more definitive term. Such an attempt by
Pochodzalla et al. through their study on $Au+Au$ reaction and analysis of
other reactions\cite{Po95} shows a behavior with characteristic of 
phase transition. Defining the temperature in terms of the yields
of He and Li, they find the temperature remains constant at about 5 MeV
for the entire range of excitation energy 3 to 10 MeV/nucleon. 
For higher excitation 
energy, the temperature increases monotonically. This has once again
brought the topic of liquid-gas phase transition in nuclear system to
the frontier of heavy-ion physics. This situation warrants theoretical
study to see if it is possible to obtain realistic caloric curve using
known features of nuclear dynamics. 

In the past, many theoretical attempts have been made to study the thermostatic
properties of hot nuclear matter using nucleon-nucleon interaction in the
framework of Thomas-Fermi models [9-11] and
temperature dependent Hatree-Fock\cite{Jq83,Ad92} models. 
Using the equation of state so obtained, critical temperatures  
in the range 15 to 20 MeV for liquid-gas phase transition in 
infinite nuclear matter have been
found. It must be emphasized here that such calculations deal with a process
in which nuclear liquid goes into nucleonic gas. This gas is supposed to have
only pure nucleons without any clusters. However, in the realistic situation,
besides the nucleons, many fragments of varying mass number will also be
produced. Hence in the theoretical calculation of caloric curve, the emission
of heavy-mass fragments need to be taken into account. A possible way for
reliable calculation may be through the statistical multifragmentation
model where the production of such fragment together with pure nucleons 
can be conveniently considered. However the key question 
is how reliably one
can calculate the excitation energy dumped into the system and the 
consequent rise in temperature.
When some energy is imparted to a nucleus there are several modes
through which the nucleus will receive the energy. 
The part of the energy which going to the 
compression or collective modes will not contribute to the rising of
temperature of the system.
Further the precise relation between the the bombarding energy 
and the excitation energy must be known in order to make contact with the  
experiments in the laboratory. The identification of the true 
mechanism of energy deposition and
consequent rise of temperature depends upon the nucleon-nucleon collisions
at a microscopic lavel, which is requires undoubtedly the solution of 
an extremely involved many-body
problem. Therefore a mechanism is invariably supposed for the calculation,
which has to be aposteriori justified through experimental support. 

Recently we have developed a model, called the dynamic statistical 
multifragmentation (DSM) model\cite{Cd97,Cd96}, for the 
intermediate energy heavy-ion 
collisions where the entrance channel characteristics like incident
energy, impact parameter and masses of colliding nuclei are taken into
account. The model is based on a spectator-participant picture and 
envisages the tripartition of the whole system into the fireball, the
projectile-like and target-like spectators. Well defined mechanisms
for the excitation of the three parts are clearly recognizable in
this model. The
excitation of the spectator parts originates from the distortion of
their shapes and that of the fireball due to the fusion of the participant
regions of the two colliding nuclei. So the excitation energy of the
fireball could be calculated using relativistic kinematics and that of
the spectators from the geometrical considerations. Then decay of the three
systems are calculated in the framework of statistical mechanics using
grand canonical picture. This model well explains the central collision 
data of $^{40}Ar$ induced $^{45}Sc$ reaction\cite{Li93},
and also the non-central collision data of $^{40}Ar$ induced $^{40}Ca$ and
$^{197}Au$ reactions\cite{Jk87}, with incident energies in the 
intermediate range of 30 to 140 MeV/nucleon. 
This success gives us the impetus
to calculate the caloric curve in the DSM model, for reactions 
which we feel will correspond to the
realistic situation and can be compared with experimental 
observations.
  
In the present work, we report our calculation of caloric curve  in 
$^{197}Au+^{197}Au$ reaction obtained by varying the incident energy
from 35 to 130 MeV/nucleon. Our notations are similar to 
Ref. \cite{Cd97,Cd96}. The spectators,
being severed from the target and projectile nuclei are  
relatively cold, and not amenable for the adequate deposition of energy 
from the projectile. So it is the fireball only, in which different
amount of energy can be deposited by varying the incident energy. 
Further, experimentally this part can be isolated kinematically from the 
spectators ones and its decay can be studied.  
So the fireball offers a convenient system to obtain the caloric
curve and study its features. For a given impact parameter we can
find\cite{Cd97,Cd96} the number of constituting nucleons in the fireball
from the geometry of the collision and the excitation energy $E^{\star}$
from the incident energy, $E_{lab}$. For different impact parameters we can 
have fireball having different mass (A) and charge (Z) numbers. 
Then we consider the 
decay of the fireball into all possible fragments of varying mass and
charge numbers detected by the available phase space in various
channels. The temperature of the fireball is determined by 
simultaneously solving the baryon number, charge number and energy 
conservation equations, as given in Ref. \cite{Cd96}. We would like to 
stress here that, in our calculation we have taken both the inter-fragment
Coulomb and nuclear interactions together into account through a 
statistical prescription[14,15,18-20].

In the DSM model, the freeze-out density of the fireball is the only 
parameter. Here we have performed our calculation with two different
densities, namely $0.22 \rho_0$ and $0.115 \rho_0$, $\rho_0$ being the
density of nuclear matter at ground-state. In Fig. 1, we have plotted the
caloric curve for the $^{197}Au+^{197}Au$ collision at three different
impact parameters, $5.8, 6.95$ and $8.8 fm$, which correspond to fireballs of 
mass and charge numbers (84, 32), (150, 60) and (196, 78) respectively. 
The upper scale
shows the incident energy of the projectile. In the figure, the
solid and the dashed lines represent the caloric curve obtained with 
two freeze-out densities $0.22 \rho_0$ and $0.115 \rho_0$ respectively. 
We find for the two heavier systems,
the temperature rises faster for very low excitation energy, upto 
$\sim$ 3 MeV/nucleon, and then the rise is slower. Between 3 to 8 
MeV/nucleon excitation energy, the temperature remains rather constant
at 5 to 6 MeV in these cases. A kink is seen in each of the four curves 
at excitation energy of about $\sim$ 8 MeV/nucleon.
Depending on the mass and the freeze-out density, the corresponding
temperatures lie within $\sim$ 6 to 6.5 MeV. 
Then with the increase in incident energy the temperature rises monotonically.  
This is comparable with the
experimental finding of Pochodzalla et al.\cite{Po95} where they 
observe the temperature 
to remain constant at 5 MeV when the excitation energy increases from 
3 to 10 MeV/nucleon, and a kink is seen at 10 meV/nucleon.
Remarkably, they characterize the density where this phenomenon is 
observed, to be in the range 
$0.15 \rho_0$ to $0.3 \rho_0$ which includes the density $0.22 \rho_0$
used in the present calculation. It may be noted that, 
in our calculation, this kink is missing in the case of lighter mass 
system $A = 84$. This suggests that in the lighter systems this 
phenomenon in not likely to be manifested. We calculated the caloric
curve for a series of systems with varying mass numbers and found
that the constancy of temperature over certain range of excitation
energy and the kink in the caloric curve starts showing up only when the
number of nucleons in the system is more than $\sim$ 120, which is in 
agreement with Gross\cite{Gr90}. However, Bondrof et al.\cite{Bd85} gets such
behavior even for low mass system $A = 100$. 
De et al. have also attempted to calculate the caloric curve for $^{150}Sm$
nucleus in Thomas-Fermi model\cite{Jd97}. However, they find a kink at a 
much higher
excitation energy of about $\sim$ 18 MeV/nucleon with a corresponding
temperature $T \sim 10$ MeV for the density $0.125 \rho_0$. They donot
find such behavior for higher density.
 
To see what effect the nature of inter-fragment interaction has on this
result, we have calculated these caloric curves with switching on and off 
the nuclear interaction which is, normally not taken into account in
many calculations\cite{Gr90,Bd85}. In Fig. 2, we have presented the 
caloric curves obtained with inter-fragment Coulomb plus nuclear interaction
and Coulomb interaction only by solid and dashed line respectively,
for the density $0.22 \rho_0$. We find, when the nuclear interaction is
switched off, the kink gets shifted to a higher value of excitation energy
of 12 MeV/nucleon with temperature of $\sim$ 8 MeV. This takes us 
substantially away from the experimental result. 
The coming down of the temperature of the fireball to the realistic value
when nuclear interaction is
included is in accord with our earlier studies. Such lowering is expected
as, the nuclear interaction being attractive in nature, tends to reduce the
kinetic energy of the fragments in the assembly and consequently the
temperature. Gross in his model study of decay of hot nuclei\cite{Gr90} 
in the framework of microcanonical formalism, finds the temperature to 
remain constant for a very short range of excitation energy.
This may be because he doesn't take the
nuclear inter-fragment interaction into account and also treats the 
neutron channel separately. However, in the present study using the 
DSM model, all the channels are treated on equal footing due to the
inclusion of inter-fragment nuclear interaction. This leads to more
realistic caloric curve with appropriate value of excitation energy 
and temperature comparable with experiment.

With a view to see whether the kink found in the caloric curve, is related
to a phase transition, we have calculated the specific heat
of the system from the caloric curve. It is the relevant observable
of the system, defined as,
\begin{equation}
C_v = {\left({d{E^{\star}}}/{dT} \right)}_v.
\end{equation}  
In Figs. 3 and 4 we have plotted the calculated $C_v$ versus the 
temperature for the density $0. 22 \rho_0$,
the fireballs of masses $A = 150$ and $196$ respectively. 
We find well defined peaked structure signaling
the possible existence of a liquid-gas phase transition 
at $T \sim 6$ MeV for $A = 150$ and $T \sim 6.5$ MeV for $A = 196$. 
This transition is expected in the nuclear system with the excitation
energy in the range 8 to 10 MeV/nucleon. 

In summary, we have obtained the caloric curve for the system $A = 84, 150$
and $196$, likely to be produced in $^{197}Au+^{197}Au$ collision. 
It is found that the mechanism of energy deposition through the 
tripartition picture of the DSM model and the inter-fragment nuclear 
interaction play the decisive role in producing a realistic caloric curve. 
The temperature is shown to remain nearly constant at 5 to 6 MeV for the 
range of excitation energy 3 to 8 MeV/nucleon, which is close to 
experimental observation.  We find such behavior is only seen when
the mass of the system is more than $\sim$ 120. A kink is seen at 
excitation energy 8 MeV/nucleon, corresponding to temperature $\sim$ 
6 to 6.5 MeV, which is speculated to be related to a liquid-gas phase 
transition. This possible signature of phase
transition is more clear from the specific heat distribution 
which shows a peak structure at this temperature. Hence this 
temperature may be treated as critical
temperature of liquid-gas phase transition in finite nuclear matter.
However, the determination of
the order of this transition and finding out proper critical exponents
are quite important factors for establishing this liquid-gas phase 
transition.

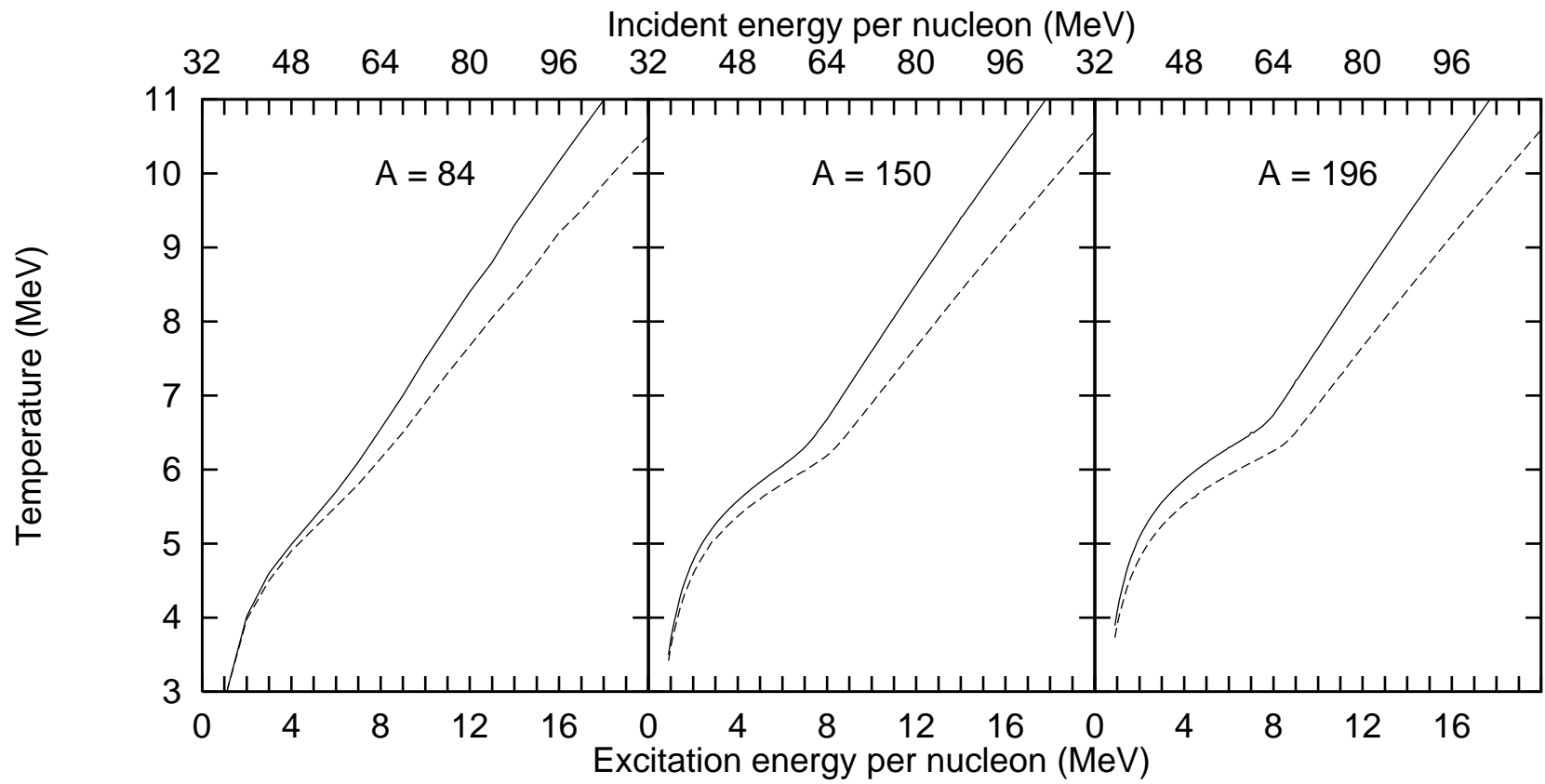
\begin{figure}
\caption{Caloric curve for the fireball with $A = 84, 150$ and $196$.
The solid and dashed lines are for densities $0.22 \rho_0$ and $0.115
\rho_0$ respectively.}
\end{figure}

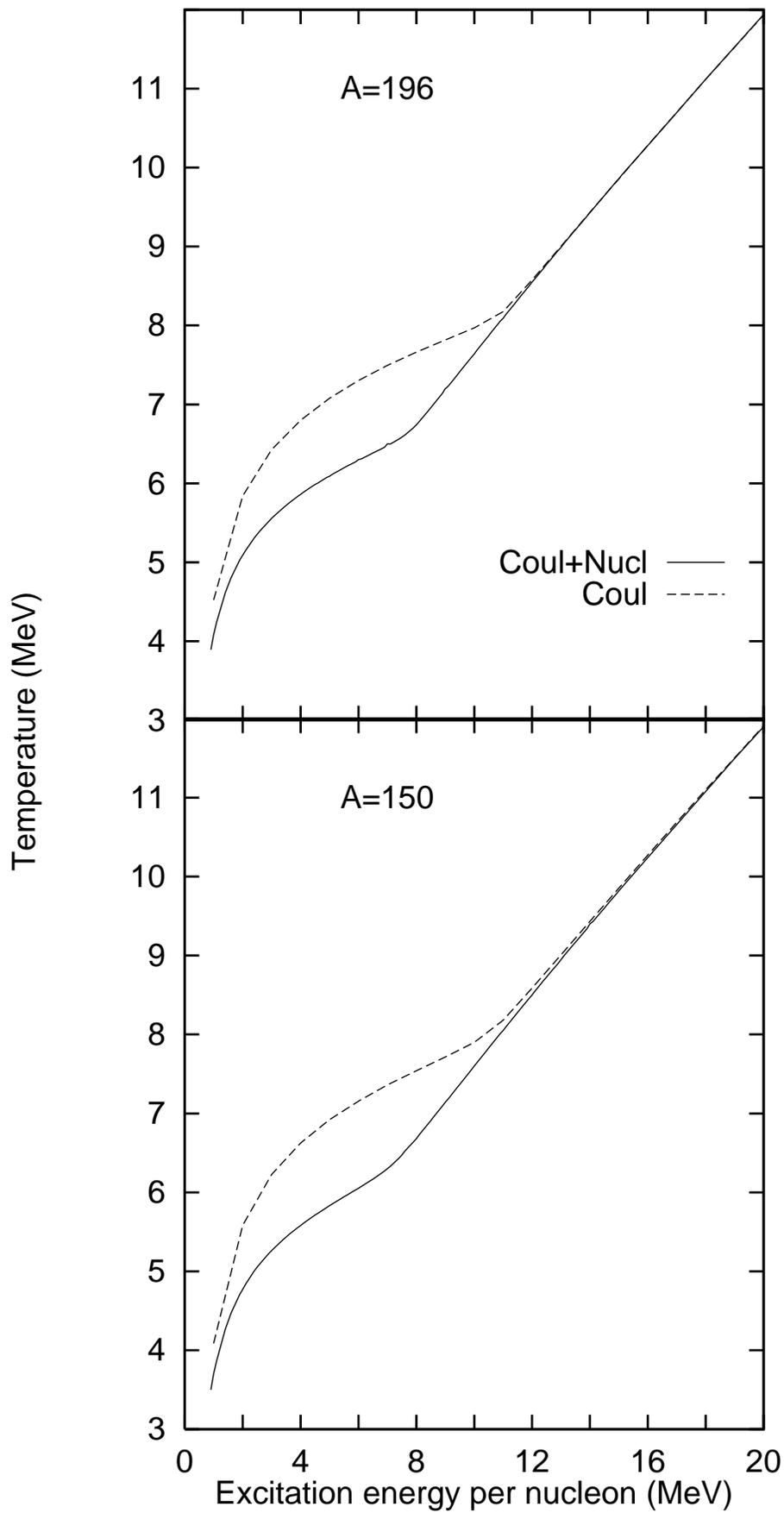
\begin{figure}
\caption{Caloric curve for the fireball with $A = 150$ and $196$. 
the solid and dashed lines are the calculation with Coulomb plus nuclear
and only Coulomb inter-fragment interaction respectively.}
\end{figure}

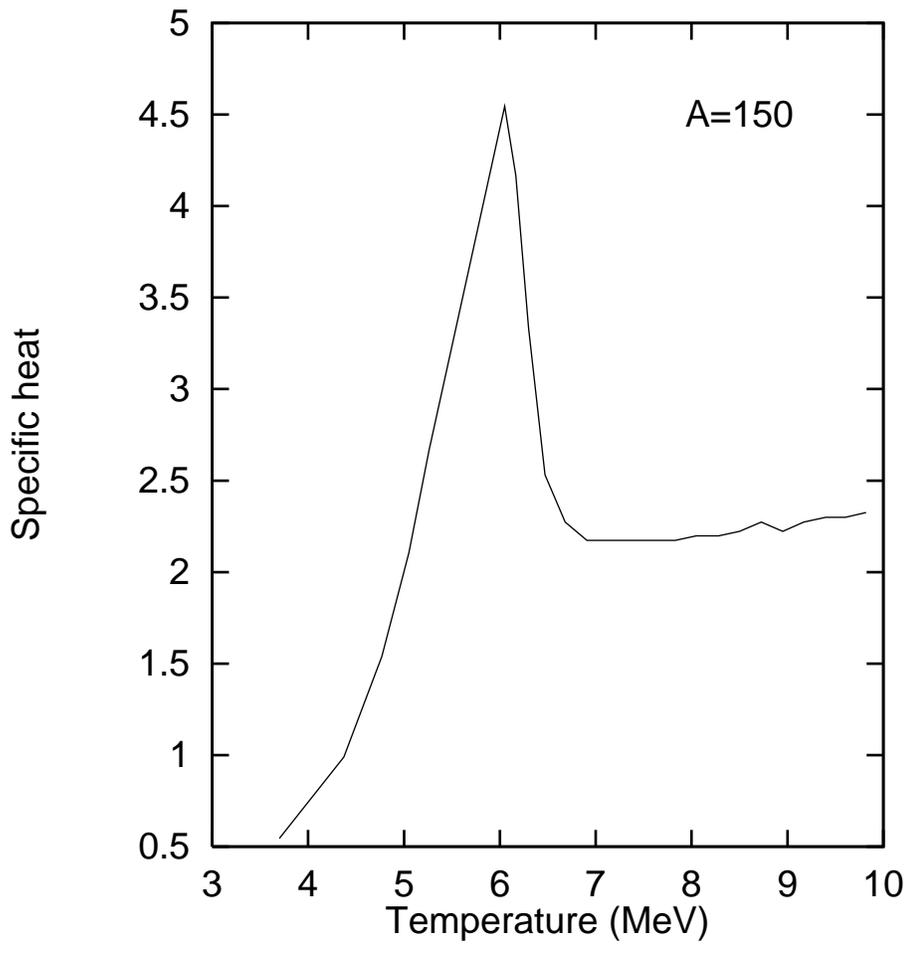
\begin{figure}
\caption{Specific heat distribution for fireball with $A = 150$.}
\end{figure}

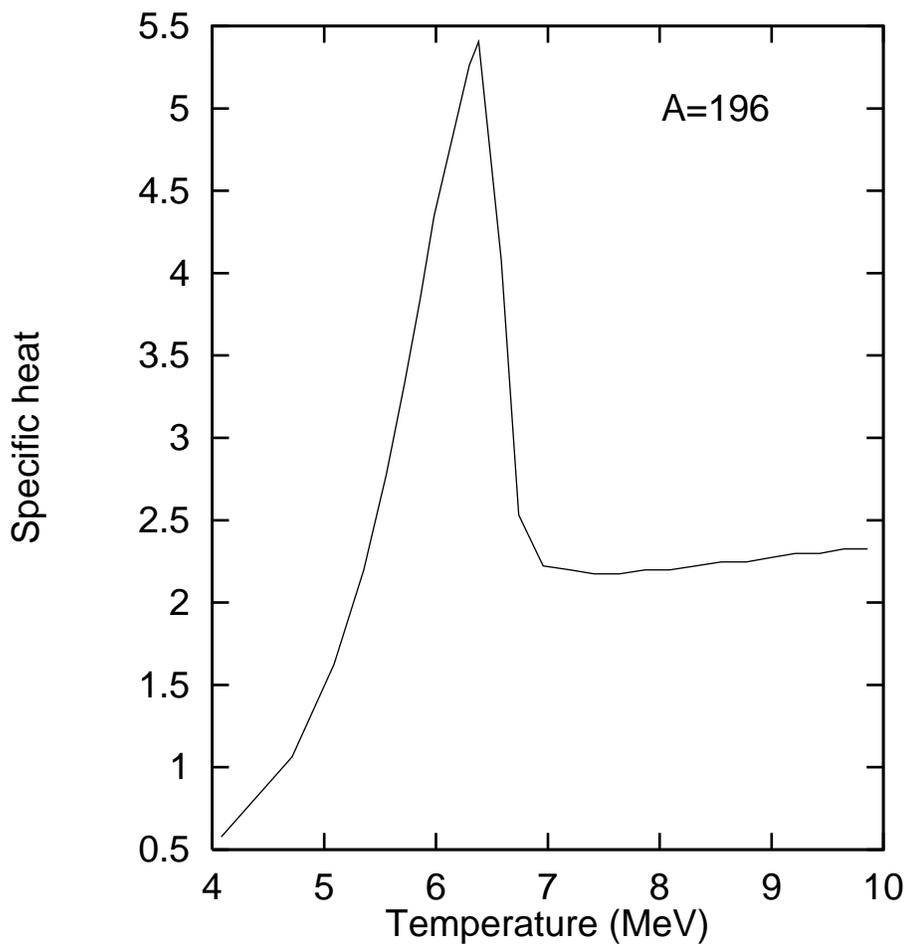
\begin{figure}
\caption{Same as Fig. 3 but for $A = 196$.}
\end{figure}

\end{document}